\documentclass[prl,twocolumn,superscriptaddress,preprintnumbers,nofootinbib]{revtex4-1}
\usepackage[subpreambles=true]{standalone}

\usepackage{graphicx}
\usepackage{dcolumn}
\usepackage{bm}
\usepackage{hyperref}
\usepackage{subfigure}
\usepackage{float}
\usepackage{wrapfig}
\usepackage{amssymb,latexsym, amsmath}
\usepackage{graphics}
\usepackage{float}
\usepackage{appendix}
\usepackage{amssymb,latexsym,amsthm,makeidx,enumerate}
\usepackage{mathrsfs}
\usepackage{siunitx}
\usepackage{comment,hhline,multirow,setspace}
\usepackage{hyperref}

\newcommand{\TaMoS}{Ta$_{1-x}$Mo$_x$S$_2$ }
\newcommand{\tas}{TaS$_2$}

\newcommand{\tasdope}{Ta$_{0.9}$Mo$_{0.1}$S$_2$}
\newcommand{\Eg}{E$^1_{2g}$}
\newcommand{\Ag}{A$_{1g}$}

\begin{document}

\makeatletter
\let\NAT@bare@aux\NAT@bare
\def\NAT@bare#1(#2){%
	\begingroup\edef\x{\endgroup
		\unexpanded{\NAT@bare@aux#1}(\@firstofone#2)}\x}
\makeatother

\makeatletter
\newcommand{\printfnsymbol}[1]{%
  \textsuperscript{\@fnsymbol{#1}}%
}
\makeatother

\title{Raman spectroscopy of few-layers \tas{} and Mo-doped \tas{} with enhanced superconductivity}

\author{S. Valencia-Ibáñez$^*$}
\author{D. Jimenez-Guerrero$^*$}
\author{J. D. Salcedo-Pimienta}
\author{K. Vega-Bustos}
\affiliation{Department of Physics, Universidad de Los Andes, Bogotá 111711, Colombia}
\author{G. C\'ardenas-Chirivi}
\affiliation{Department of Physics, Universidad de Los Andes, Bogotá 111711, Colombia}
\affiliation{Facultad de Ingeniería y Ciencias Básicas, Universidad Central, Bogotá 110311, Colombia}
\author{L. López-Manrique}
\affiliation{Department of Chemistry, Universidad de Los Andes, Bogotá 111711, Colombia}
\author{J. A. Galvis}
\affiliation{Facultad de Ingeniería y Ciencias Básicas, Universidad Central, Bogotá 110311, Colombia}
\author{Y. Hernandez}
\author{P. \surname{Giraldo-Gallo}$^{\dagger}$}
\affiliation{Department of Physics, Universidad de Los Andes, Bogotá 111711, Colombia}

\date{\today}

\begin{abstract}

The use of simple, fast and economic experimental tools to characterize low-dimensional materials is an important step in the process of democratizing the use of such materials in laboratories around the world. Raman spectroscopy has arisen as a way of indirectly determining the thickness of nanolayers of transition metal dichalcogenides (TMDs), avoiding the use of more expensive tools such as atomic force microscopy, and it is therefore a widely used technique in the study of semiconducting TMDs. However, the study of many metallic TMDs in the limit of few atomic layers is still behind when compared to their semiconducting counterparts, partly due to the lack of similar alternative characterization studies. In this work we present the characterization of the Raman spectrum, specifically of the \Eg{}- and \Ag{}-modes, of mechanically exfoliated crystals of \TaMoS, a metallic TMD which exhibits charge density wave formation and superconductivity. The clear identification of contributions to the Raman spectrum coming from the SiO$_2$/Si substrate, which overlap with the peaks coming from the sample, and which dominate in intensity in the few-layer-samples limit, allowed the isolation of the individual \Eg{}- and \Ag{}-modes of the samples and, for the first time, the observation of a clear evolution of the Raman shifts of both modes as a function of sample thickness. The evolution of such peaks qualitatively resembles the evolution seen in other TMDs, and provide a way of indirectly determining sample thickness in the limit of few atomic layers at a low cost. In addition, we observe a softening (red-shift) of both \Eg{}- and \Ag{}-modes with Mo-doping in the nanolayers, possibly related to the increased out-of-plane lattice parameter with respect to the pure compound.
\end{abstract}

\maketitle

\def\thefootnote{*}\footnotetext{These authors contributed equally to this work}\def\thefootnote{\arabic{footnote}}
\def\thefootnote{$\dagger$}\footnotetext{Corresponding author: pl.giraldo@uniandes.edu.co}\def\thefootnote{\arabic{footnote}}

%
\vspace{2pc}

\section{Introduction}

\begin{figure}[!t]
\vspace{0.3cm}
\hspace{-0.3cm}
\centering
\includegraphics[width = \linewidth]{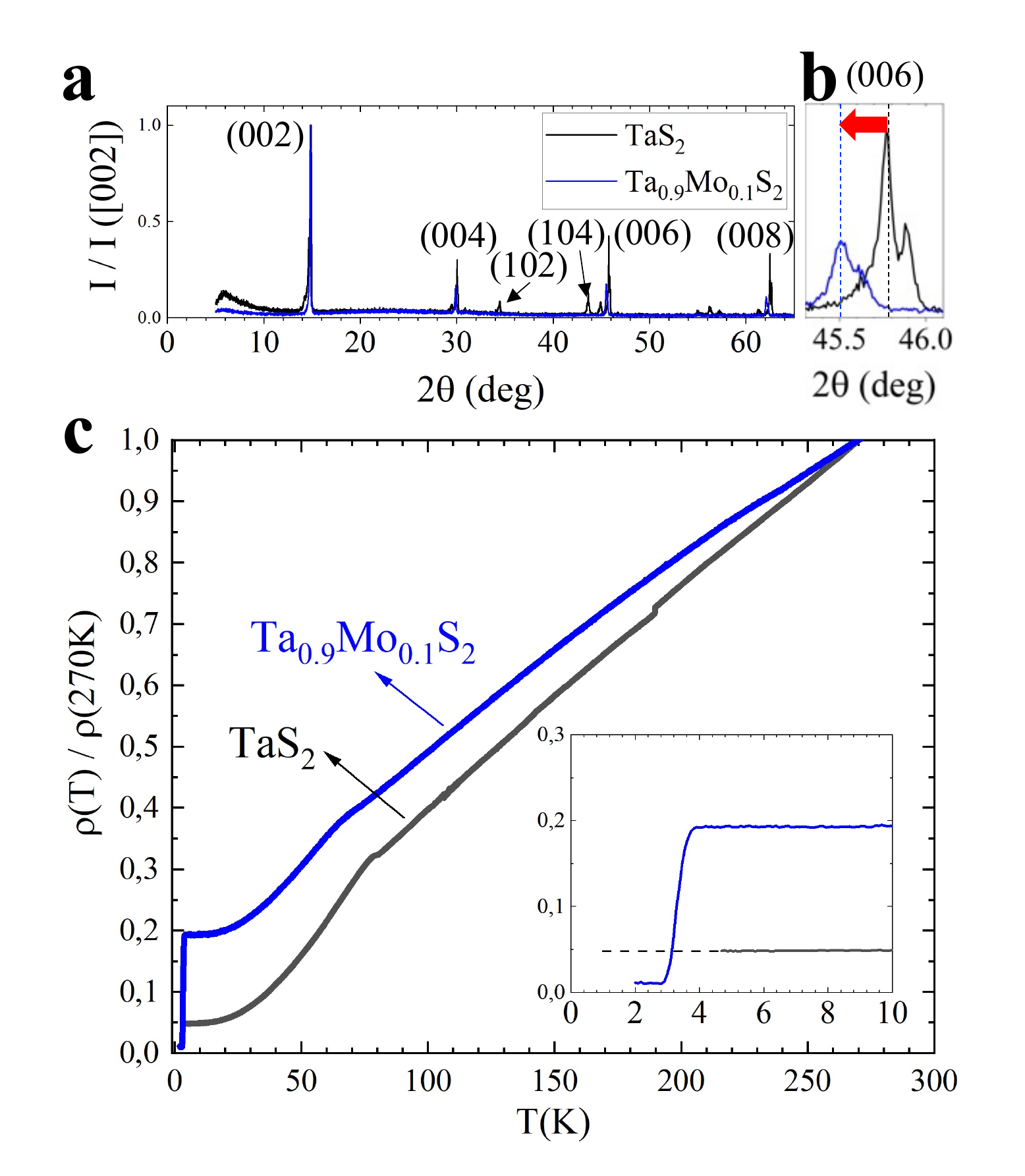} 
\vspace{0.2cm}
\caption{(Color online) \textbf{(a,b)} X-ray diffraction data for a collection of single crystals of TaS$_2$ (black curve) and Ta$_{0.9}$Mo$_{0.1}$S$_2$ (blue curve), taken in an Eulerian-cradle geometry. For both types of samples, diffraction peaks are consistent with a 2H-phase, for which the peak labeling was done. For the doped-compound the lattice parameter in the $c$ direction is increased by 0.5$\%$, as evidenced in the shift of the $\{00n\}$ peaks to lower angles, and better appreciated in the close-up look to the (006) peak in (b). \textbf{(c)} In-plane resistivity as a function of temperature, normalized by their values at 270 K, for representative TaS$_2$ and Ta$_{0.9}$Mo$_{0.1}$S$_2$ bulk single crystals. For the pure bulk-compound, superconductivity has been reported only below $\sim$0.5 K \cite{Abdel-Hafiez2016} (out of our range of temperatures). The Mo-doped samples show superconductivity at a temperature close to 4 K (onset), as better seen in the inset to the figure. The CDW transition temperature, seen at the point where the resistivity curves go from approximately linear to approximately quadratic (when cooling) is also slightly decreased with Mo-doping.
}\label{fig1}
\end{figure}

\begin{figure}[!t]
\vspace{0.3cm}
\hspace{-0.3cm}
\centering
\includegraphics[width = \linewidth]{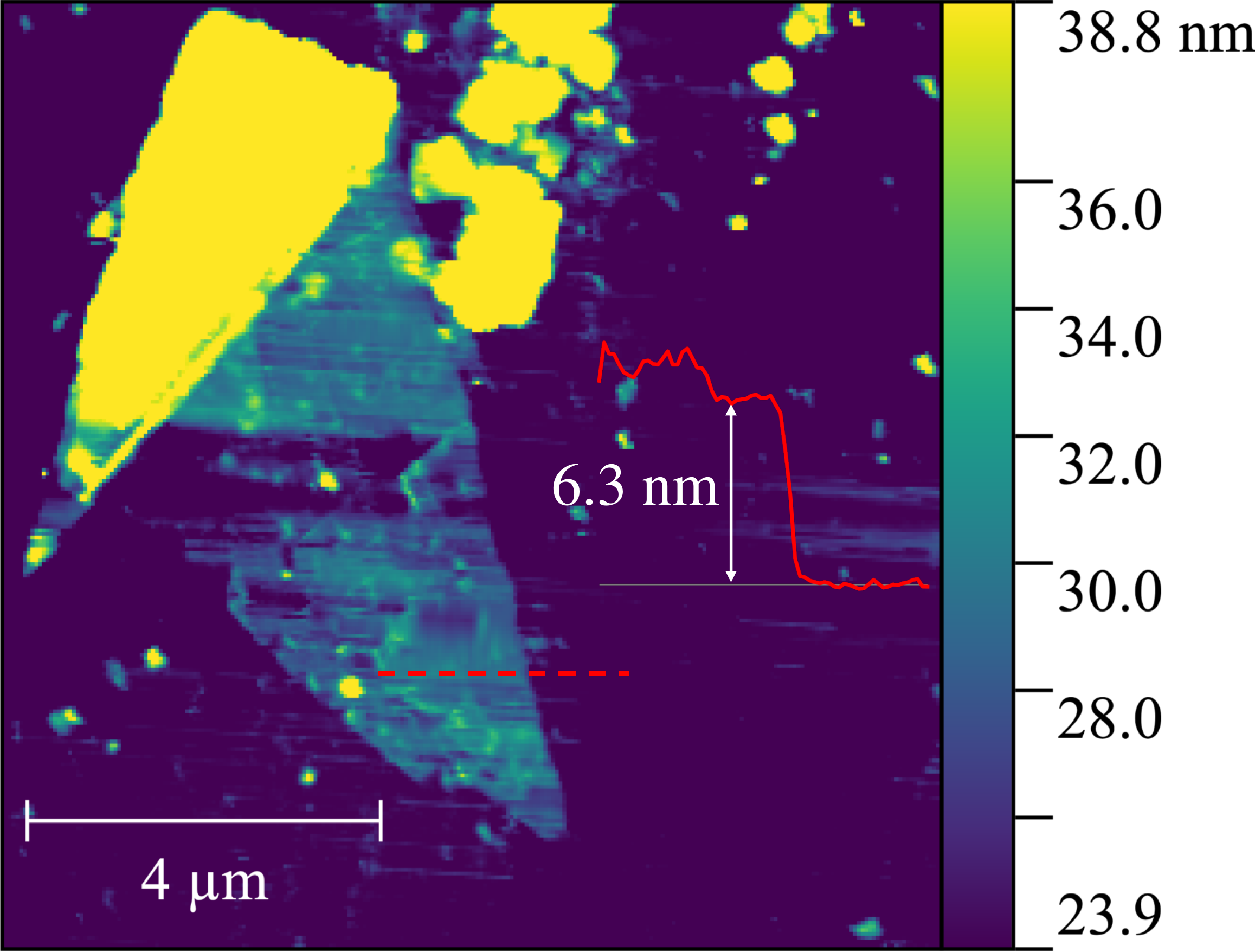} 
\vspace{0.2cm}
\caption{(Color online) AFM topographic image of a representative exfoliated TaS$_2$ crystal deposited on a Si/SiO$_2$ substrate. The color scale represents the relative height, as indicated in the color bar. A topographic profile of the path shown by the dashed red line is inserted. The height of the specified section is approximately 10 atomic layers (5 unit cells). This same procedure was followed for determining the number of layers of all samples measured in this work.
}\label{fig2}
\end{figure}

Transition metal dichalcogenides (TMDs) are among the most versatile families of materials. These are layered materials with general chemical formula MX$_2$, where M is a transition metal and X a chalcogen atom. Atoms in this structure are strongly-bonded in the  \textit{a-b} plane but weakly-bonded and stacked by van der Waals forces in the \textit{c} direction, which gives them their low-dimensional character \cite{2016Castellanos,2017Manzeli,CHOI2017,Chowdhury2020,2021Xinmao}. The weak van-der-Waals stacking of the planes allows the isolation of single- or few-atomic-layers-thick samples through relatively simple mechanical exfoliation techniques \cite{Zhao_2020, Gant_2020}. One of the most outstanding characteristics of this family of materials is that, depending on chemical composition, their electronic properties can range from band insulators to metals, and they can show exotic quantum phenomena such as charge density waves (CDWs), magnetic ordering, superconductivity, topological electronic states, among others \cite{kuc_heine_kis_2015,Kuc2015,2017Manzeli}. Among all the members of this family of compounds, the semiconducting members are one the most studied materials in the last two decades. Their visible or near-visible range bandgaps, elevated electronic mobilities, large spin-orbit interaction, large susceptibility to changes in physical parameters, spatial scalability, among other characteristics, makes them ideal candidates for a large number of optoelectronics, spintronics, and electronics applications \cite{Rold_n_2015,2016Castellanos, 2017Manzeli, 2017Duong,Chaves2020}. Interestingly, all of these characteristics get enhanced or only appear in the limit of few atomic layers, and the body of literature characterizing electronic and physical properties in the monolayer limit is extensive \cite{Splendiani2010,LI2017}. 

\begin{figure*}[!t]
\vspace{0.3cm}
\hspace{-0.3cm}
\centering
\includegraphics[width = \linewidth]{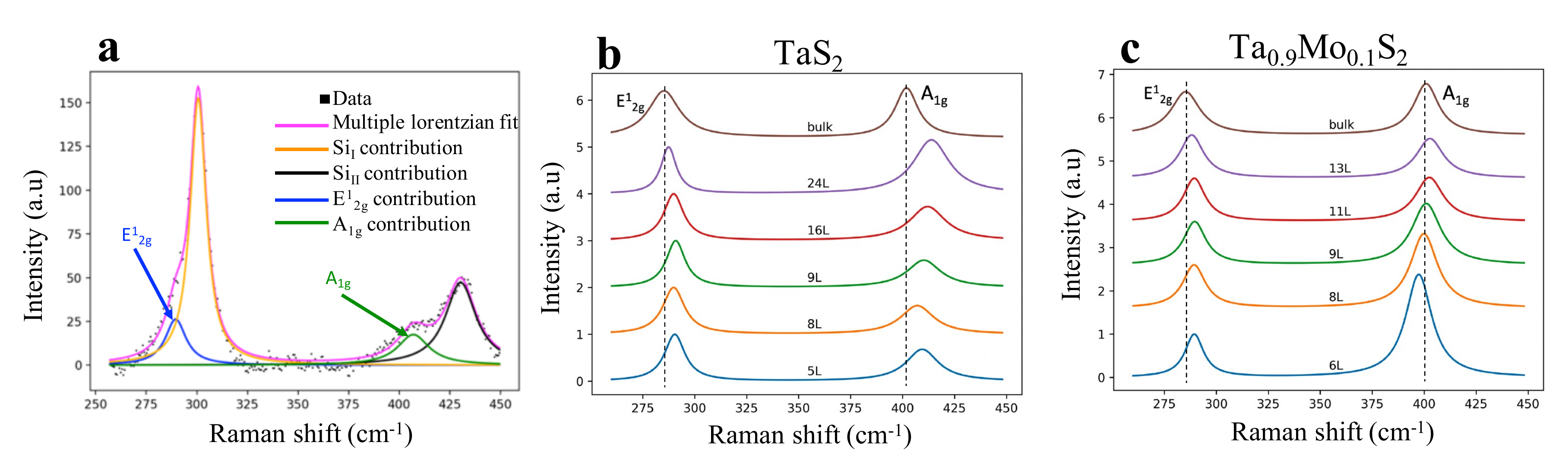} 
\vspace{0.2cm}
\caption{(Color online) \textbf{(a)} Characteristic Raman spectra of a representative exfoliated sample of TaS$_2$ with a low number of layers, deposited on a Si/SiO$_2$ substrate. The spectrum is shown only in a frequency range around the E$^1_{2g}$ and A$_{1g}$ peaks. The background of the spectra has been subtracted to highlight only the peaks. Four different contributions to the spectra can be seen: the most intense peaks are associated to vibrational modes of the Si/SiO$_2$ substrate (and are fixed in frequency for all samples measured), and the weaker peaks are the ones associated to the \Eg{}- and \Ag{}-modes of the exfoliated sample. Lorentzian fits are shown for all peaks, from which the central frequency of the modes are extracted. \textbf{(b)} Raman spectra of TaS$_2$ and \textbf{(c)} Ta$_{0.9}$Mo$_{0.1}$S$_2$ samples with different number of layers, after subtracting the substrate contribution. The vertical dashed-line marks the central frequency of the \Eg{} and \Ag{}-modes in the bulk limit. A clear shift of the central frequency of the E$^1_{2g}$ and A$_{1g}$-modes is observed between the few-layered and bulk crystals.
}\label{fig_RamanSpectra}
\end{figure*}

On the other hand, the availability of characterizations of the properties of \textit{metallic} TMDs in the limit of few atomic layers is more scarce. Recent works have reported the tunability of the electronic properties of metallic TMDs such as NbSe$_2$ \cite{2015Xi,2016Xi,2017Xing,2018DeLaBarrera,2019Zhao}, TaSe$_2$ \cite{2013Galvis,2018Ryu,2018Wu,2019Lian} and \tas{} \cite{2014Galvis,2016Navarro-Moratalla, Abdel-Hafiez2016, 2018DeLaBarrera} in the limit of few atomic layers, which, just as with their semiconductor counterparts, indicates an optimization of the novel properties of these compounds in the pure 2D limit. 
Among these, the 2H polytype of \tas{} stands out for having one of the highest reported superconducting transition temperatures ($T_c$) of all TMDs (9.15 K at a pressure of 8.7 GPa \cite{Abdel-Hafiez2016}). At ambient pressure, bulk samples of \tas{} exhibit an incommensurate CDW below 76 K and superconductivity below $\sim$0.5 K. Various avenues for enhancing the superconducting critical temperature (at atmospheric pressure) have been explored, such as chemical substitution with Se \cite{2017Li}; intercalation by Cu \cite{2008Wagner}, Pd \cite{Zhou_2018}, Ni \cite{2010Li}  and Na \cite{2005Fang}; as well as dimensional confinement by mechanical exfoliation down to few-layers thick samples, for which $T_c$ can go up to 3 K in pure \tas{} monolayers \cite{2014Galvis,2016Navarro-Moratalla, 2018DeLaBarrera}, or 3.6 K in \tas{} monolayers with structural defects \cite{2018Peng}. This is, about seven times larger than in the bulk. 

These exciting observations, particularly for \tas{} in the monolayer limit, invite to find better, faster and cheaper ways of producing few-layer samples as well as determining their thickness and properties. Consequently, Raman spectroscopy has risen as a fast and trustworthy way to indirectly determine the thickness of TMD nanolayers \cite{2012Li,2013Castellanos,Liang_2018}, and as an alternative to more direct but expensive techniques for thickness determination such as atomic force microscopy (AFM). Raman spectroscopy characterizations are widely available for semiconducting TMDs, which has helped democratizing the study of these materials, and as a consequence, advancing in the understanding of their properties in the pure 2D limit and their integration into devices. This has not been the case for metallic TMDs, and it is therefore necessary to perform such characterizations in order to allow easy and fast identification of ultrathin samples of, not only the pure metallic compounds, but also the chemically doped ones for which the novel ground states are improved.

In this paper we report the characterization of the Raman shift of the \Eg{} and \Ag{} vibrational modes of 2H-\tas{} and 2H-\tasdope{} single crystals as a function of the number of atomic layers. For both compounds we find, for the first time, a clear but distinct evolution of the \Eg{} and \Ag{} Raman peaks as a function of the number of layers, in contrast with previously published works in pure \tas{} for which peaks associated with the substrate in which samples are deposited were not considered \cite{2016Navarro-Moratalla, 2016Island}. In addition, we observe a softening (red-shift) of both \Eg{} and \Ag{}-modes with Mo-doping in the limit of few-atomic layers. This red-shift is lost in the bulk-limit, for which the Raman-shift of the \Eg{} and \Ag{}-modes are the same in the pure and doped compound.

\section{Methods}
\subsection{Single-crystal synthesis}
TaS$_2$ and Ta$_{0.9}$Mo$_{0.1}$S$_2$ bulk single crystals were synthesized in a two-step process. First, the stoichiometric ratios of Ta, S and Mo, were thoroughly ground and pressed into pellets. The pellets were sealed in vacuum inside a quartz tube and sintered at 900 $^\circ$C for 100 hours. The resulting polycrystalline compounds were ground again and sealed with 5 mg/cc iodine in a long evacuated quartz tube. Finally, single-crystals were obtained by chemical vaport transport (CVT). The tube was placed inside a horizontal one-zone tube furnace for 7 days, with a center-of-furnace temperature of 1000 $^\circ$C, at which the powder-containing end of the quartz-tube was placed. The other end of the tube was placed toward one end of the furnace, at which a temperature difference of the order of 100$^\circ$C with respect to the center of the tube furnace is naturally developed. Shiny-platelet-like crystals were obtained in the cold end of the tube, with typical lateral sizes of 1 mm $\times$ 1 mm. 

\subsection{Few-layered crystals exfoliation}
Few-layer crystals of TaS$_2$ and Ta$_{0.9}$Mo$_{0.1}$S$_2$ were obtained through repeated mechanical exfoliation using a combination of clean Nitto SPV 224 tape for the initial exfoliations, and Gel-pack X4 viscoelastic poly-dimethylsiloxane (PDMS) stamps for the final exfoliations. The PDMS-stamps were then pressed against 300 nm SiO$_2$/Si substrates to deposit a collection of exfoliated crystals.

\subsection{Characterization}
The structural characterization of the obtained crystals was performed using X-ray diffraction in a Eulerian-Cradle geometry, in a Panalytical X-pert diffractometer. This characterization was performed in a collection of single crystals of each batch, which were cut in tiny fragments and spread in double-sticky tape to emulate a powder diffraction experiment. This was done with the purpose of obtaining a statistically meaningful characterization of many crystals of each batch from which each individual crystal for the exfoliation process is taken.\\
The precise thickness of the exfoliated \TaMoS{} flakes, deposited on the SiO$_2$/Si substrates, was determined using atomic force microscopy (AFM). An Asylum Research MFP-3D-BIO microscope was operated under ambient conditions in contact mode. The Raman-shifts of the samples measured in AFM were then determined in a HORIBA Scientific XPLORA Xl041210 Raman spectrometer. The excitation wavelength used for all measurements was 532 nm with a grating of 2400 lines/mm and an acquisition time of 10 s.\\
In addition, resistivity measurements from room temperature down to $\sim$2 K were done in bulk single crystals, using a four-probe technique. This allowed identifying the critical temperatures of the CDW and superconducting transitions.

\begin{figure*}[!t]
\hspace{-0.3cm}
\centering
\includegraphics[width = \linewidth]{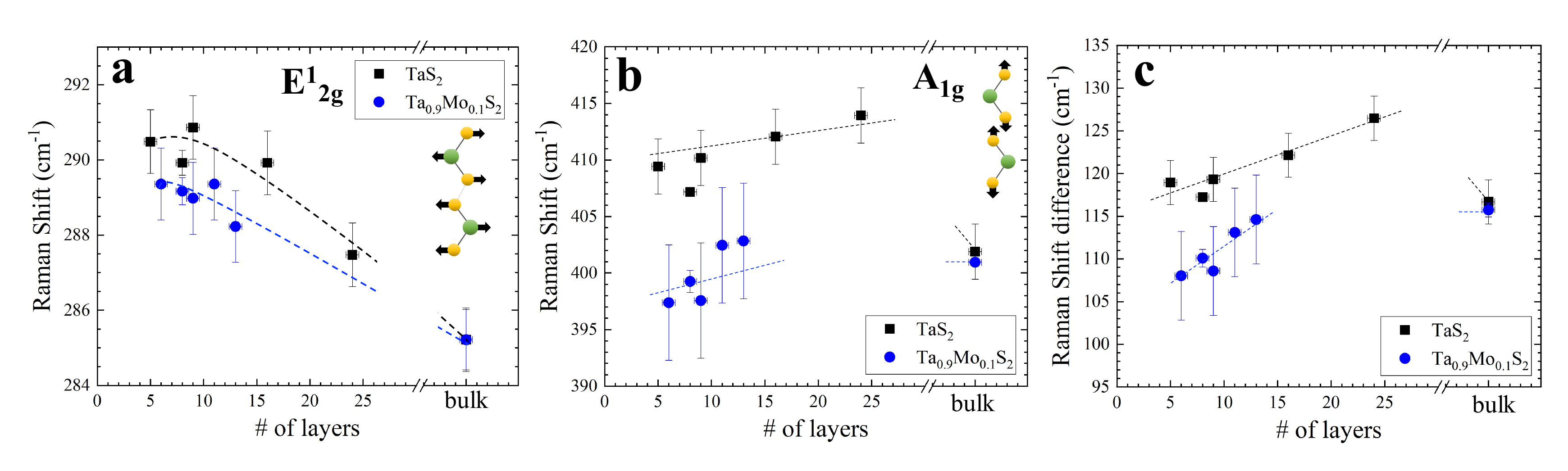} 
\caption{(Color online) Raman shift frequency as a function of the number of layers for  TaS$_2$ and Ta$_{0.9}$Mo$_{0.1}$S$_2$ samples, for the \textbf{(a)} E$^1_{2g}$- and \textbf{(b)} A$_{1g}$-modes. Error bars represent the dispersion among two to three spectra taken in different points of the sample, for each layer number. Insets to both figures show an schematic of the atomic displacements for both modes, with yellow balls representing S atoms, and green balls representing Ta or Mo atoms. \textbf{(c)} Difference of the Raman shifts of the \Ag{} and \Eg{}-modes, as a function of the number of layers for  TaS$_2$ and Ta$_{0.9}$Mo$_{0.1}$S$_2$ samples. For all figures dashed lines are a guide to the eye.
}\label{fig_RamanShift}
\end{figure*}

\section {Results and discussion}
X-ray characterization of our single crystals reveal that the polytype of the  TaS$_2$ and Ta$_{0.9}$Mo$_{0.1}$S$_2$ single crystals is 2H (figure 1a,b). The observed diffraction peaks correspond mainly to the $\{001\}$ family of planes. Peaks corresponding to other families of planes have very small intensity, due to the preferential orientation of the small pieces of single crystals with their a-b planes parallel to the mount surface. The data reveals a 0.5$\%$ increase in the out-of-plane $c$ lattice parameter for the Mo-doped samples, with respect to the pure \tas{} samples, evidenced in the shift to lower angles of the $\{00n\}$ diffraction peaks. This shift can be better appreciated in the close-up look to the (006) peak in figure 1b. 

Resistivity measurements as a function of temperature were taken in selected bulk single-crystals of TaS$_2$ and Ta$_{0.9}$Mo$_{0.1}$S$_2$, and are shown in figure 1c. These curves confirm the crystallization of the samples in the 2H-phase, as the curves are characteristic of samples of this polytype: both compositions show a transition to a CDW state below about 80 K for the pure TaS$_2$ samples, and 75 K for the Mo-doped samples. In addition, the Mo-doped sample shows a superconducting transition with onset of about 4 K. This is, about 10 times higher than the value reported for pure TaS$_2$, and it is the highest for all values of Mo-doping in \tas{} \cite{2021Salcedo}.

For both types of samples, \tas{} and \tasdope{}, few-layer crystals were mechanically exfoliated from single-crystal bulk samples and deposited on SiO$_2$/Si substrates (see methods). Inspection of the samples under the optical microscope enables a quick identification of possible few-layered crystals due to differences in interference colors \cite{2004Kvavle,2013Castellanos}. The selected flakes were characterized using both atomic force microscopy (AFM) and Raman spectroscopy. Figure \ref{fig2} shows an AFM topographic image of a representative exfoliated \tas{} sample, in which a large area of approximately 10 layers thick (6.3 nm) is identified. Samples of different heights were identified by AFM measurements, with typical lateral sizes of a couple microns. These same samples were afterwards characterized through Raman spectroscopy.

The raw Raman spectra were processed to obtain the contributions of the \Eg{} and \Ag{} modes. First, the fluorescence baseline was removed using an asymmetric least squares smoothing algorithm \cite{2005Eilers}. Figure \ref{fig_RamanSpectra}a shows a typical Raman spectrum (after subtracting the fluorescence background) of few-layered 2H-\tas{} flakes. As our interest is the characterization of the \Eg{} and \Ag{} modes, we will focus on Raman shifts around these peaks: 285 cm$^{-1}$ and 400 cm$^{-1}$ in the bulk, respectively, which matches previously reported values \cite{Sugai1981,1983Hangyo,1985Sugai,2020Zhang}. For bulk samples, only these two peaks are observed in the baseline-corrected spectra for the range of Raman shifts from 250 cm$^{-1}$ to 450 cm$^{-1}$. However, for few-layered samples, two additional peaks, with a growing relative intensity as sample thickness decreases, are identified in this Raman shift range. The positions of these peaks, which dominate the spectra for the thinnest samples, is fixed for all samples measured, and matches the peaks found in measurements of the substrate alone under the same measurement conditions. These substrate peaks match the ones reported by previous works in SiO$_2$/Si: the one at 300 cm$^{-1}$ is reported to be originated in the Si, and the one at 428 cm$^{-1}$ has been reported to come from amorphous SiO$_2$ on SiO$_2$/Si substrates \cite{Dzhurkov2014,2020Hruby}. As mentioned, the Raman signal for the thinnest samples is dominated by the substrate peaks, which almost overlap the weak \Eg{} and \Ag{} sample modes. This is the reason why previous works have failed to detect a marked evolution of the Raman shift of the \Eg{} and \Ag{} modes as a function of the number of layers in \tas{} \cite{2016Navarro-Moratalla, 2016Island, 2020Zhang}. The overlap between Raman modes frequencies for substrate and sample makes the analysis of the Raman spectrum of nanolayers of \tas{} particularly challenging, in contrast with other dichalcogenides in which this effect is not as important. In order to isolate the individual contributions from each vibrational mode in \tas{} samples, a four-peak Lorentzian least-squares fitting was used. Figure \ref{fig_RamanSpectra}a shows the shape of the four Lorentzians found from the fit, for a representative thin \tas{} sample, from which the central frequencies of the \Eg{} and \Ag{} modes are extracted.

Figures \ref{fig_RamanSpectra}b and \ref{fig_RamanSpectra}c display the \Eg{} and \Ag{} modes contributions to the Raman spectra of representative \tas{} and \tasdope{} samples, for various numbers of layers, after subtracting the fluorescence background signal and neglecting the contributions from the substrate. The intensities are normalized such that the \Eg{} peak has an intensity of 1 unit for each curve. For both \tas{} and \tasdope{} samples, an evolution of the central frequency is observed for both peaks as the number of layers are decreased from bulk down to 5/6 layers (thinnest samples that we could measure using both AFM and Raman for the same sample). Figure \ref{fig_RamanShift} summarizes the evolution of the Raman shifts of the \Eg{} and \Ag{} peaks as a function of the number of layers for the pure and doped \tas{} compounds. The evolution of the Raman shift of the \Eg{} mode with number of layers, shown in figure \ref{fig_RamanShift}a, follows the same behavior than other transition metal dichalcogenides such as MoS$_2$ \cite{2010Lee,2012Li}, WS$_2$ \cite{2013Berkdemir,Zeng2013}, WSe$_2$ \cite{Zeng2013} and TaSe$_2$ \cite{2013Castellanos,Hajiyev2013}, for both, the pure and doped compounds. This softening (red-shift) of the mode with increasing number of layers has been previously interpreted in terms of an enhancement of the dielectric screening of the long-range Coulomb interactions with each added layer, which conversely leads to a weaker effective force constant between layers, and therefore, a lower frequency of oscillation of the mode \cite{2010Lee,Liang_2018,Hajiyev2013}. Noteworthy, even though the frequency of the \Eg{} mode in the bulk is the same for both the pure and doped compounds, they differ by a measurable amount for a low number of layers. The reduced frequency for the doped compound for low number of layers can be related to the increased $c$ lattice parameter revealed by our X-ray diffraction data, suggesting a larger average distance between atoms and therefore a reduced restoring force acting on them. Our data suggests that this effect is enhanced for a low number of layers, whereas it is negligible in the bulk, where the \Eg{} mode frequency coincides for both, the pure and doped compounds.

The evolution of the Raman shift of the \Ag{}-mode with number of layers is shown in figure \ref{fig_RamanShift}b. For low number of layers, the evolution of this mode follows the trend observed for other dichalcogenides: a stiffening of the mode with increasing number of layers \cite{2010Lee,2012Li,2013Berkdemir,Hajiyev2013}. Such blue-shift can be understood within a model of classical harmonic oscillators coupled through van der Waals forces, as a greater number of layers increases the van der Waals interaction between layers, which in turn increases the oscillators' effective spring constant \cite{2010Lee,Liang_2018,Hajiyev2013}. This model can also account for the lower \Ag{}-mode frequency values in the doped compound, compared to the pure compound, as the increased out-of-plane lattice parameter results in a lower effective spring-constant. This change in out-of-plane lattice parameter between the pure and doped samples has a larger effect in the \Ag{}-mode than in the \Eg{}-mode, evidenced in a frequency difference (for the lowest number of layers) of $\sim$ 12 cm$^{-1}$ versus $\sim$ 1 cm$^{-1}$, respectively. This difference is a consequence of the symmetry of the modes: the \Ag{}-mode is driven by out-of-plane atomic displacements (see inset to figure \ref{fig_RamanShift}b), whereas the \Eg{}-mode is driven by in-plane atomic displacements (see inset to figure \ref{fig_RamanShift}a), therefore implying that a change in out-of-plane lattice parameter will have a stronger influence in the \Ag{}-mode. Another important point to highlight is that the stiffening of the \Ag{}-mode with increasing number of layers is maintained up to bulk-thicknesses for the Mo-doped compound. However, for the pure \tas{} the \Ag{}-mode softens in the bulk with respect to the the nanolayered regime. This is a unique feature not previously reported for other dichalcogenides.

The difference in Raman shifts of the \Ag{} and \Eg{} peaks as a function of the number of layers is plotted in figure \ref{fig_RamanShift}c. This difference is widely used as a more sensitive indication of the number of layers for semiconducting TMDs, such as MoS$_2$ \cite{Hajiyev2013} and WSe$_2$ \cite{Zeng2013}, being minimum for the monolayer and increasing for increasing number of layers. The same trend is followed for pure and Mo-doped \tas{} in the limit of few layers. This plot can serve as a guideline for the indirect determination of the thickness in nanolayers of the \tas{} system.

\section{Conclusions}

The characterization of the Raman spectra in \tas{} and its evolution with number of layers has been a challenging task, as its most characteristic Raman peaks overlap with substrate peaks, being these last the ones that dominate the Raman signal in the limit of a few-atomic layers. In this work we were able to reveal, for the first time, a clear evolution of the \Eg{} and \Ag{} Raman peaks as a function of the number of layers in 2H-\tas{} and 2H-\tasdope{} by explicitly fitting Lorentzian functions to the peaks of both substrate and samples, for samples with different thicknesses. We find that the evolution of the \Eg{}-mode with increasing number of layers for both the pure and doped compounds, follows the same softening observed for other prototypical transition metal dichalcogenides. The \Ag{}-mode shows a stiffening as a function of increasing number of layer for the \tasdope{} samples, and equivalently for the pure-\tas{} sample in the limit of few layers. This observation analogous to what has been previously observed for other dichalcogenides. However, for the pure \tas{}, the \Ag{}-mode frequency shows a significant reduction with respect to the values in the nanolayer limit, a feature not previously reported for any dichalcogenide. Nevertheless, in the limit of few atomic layers, the behavior of the \Eg{} and \Ag{} Raman peaks in 2H-\tas{} and 2H-\tasdope{} resemble what has been reported for other dichalcogenides, which allows using the Raman-shift of these modes, as well as their difference, as an affordable alternative for inferring the sample thickness.  

We find that the effect of Mo-doping in the Raman shift of the \Eg{} and \Ag{} Raman peaks, in the limit of a few atomic layers, is the softening of both modes. This effect is ten times larger for the \Ag{}-mode than for the \Eg{}-mode. This softening could be explained by the increase in out-of-plane lattice parameter induced by Mo-doping, which leads to a decreased van der Waals interaction between layers, and therefore to a reduced force constant and oscillation frequency of the modes. Interestingly, the effect of Mo-doping on the \Eg{} and \Ag{} \textit{in the bulk-limit} is negligible, as the Raman shifts of both modes are the same (or at least, within the error bars) for the pure and doped compounds in this limit. A possible explanation for this could be a larger relative effect of the lattice parameter change with respect to the sample thickness, which can lead to an enhancement of this effect for the ultra-thin samples; or alternatively, a larger lattice parameter increase for the Mo-doped samples as thickness is reduced.

\section*{Acknowledgments}

We thanks Ian R. Fisher and Joshua Straquadine for support during the sample growth process, and insightful discussions. 
S.V-I., D.J-G., J.D.S-P., J.A.G., and P.G-G. thank the support of the Ministerio de Ciencia, Tecnología e Innovación de Colombia (Grants No. 120480863414 and 122585271058). J.D.S-P., L.L-M, P.G-G. and Y.H. thank the support of the School of Sciences of Universidad de Los Andes. P.G-G. thanks the support of the FAPA grant from Vicerrectoría de Investigaciones, Universidad de Los Andes, and  the L'Oréal-UNESCO For Women in Science International Rising Talents Programme. The authors thank the Microcore Center for Microscopy of Universidad de los Andes.

	
\bibliography{1_TaS2RamanPaper_ArXiv.bbl}\vspace{-0.8cm}

\end{document}